# The influence of gain compression factor on dynamical properties of single level InAs/GaAs quantum dot lasers

**Mostafa Qorbani, Esfandiar Rajaei[*], Omid Hajizadeh and Mahdi Ahmadi Borji**

Department of Physics, University of Guilan, Rasht, Iran

[*]Corresponding author, Email: Raf404@guilan.ac.ir

**Abstract**

In this paper, by representing a single level rate equation model for InAs/GaAs quantum dot lasers and computations by fourth order Runge-Kutta method some characteristics of the output laser are considered. The change of photon number in time and current and also the output power versus current with different gain compression factors are investigated for lasing from ground state (GS). Afterwards, the response function of small signal modulation for ground state in constant current but different gain compressions is surveyed. At last, we find an optimum value for gain compression factor in lasing from GS.

## I. Introduction

Semiconductor lasers have found a big application in telecommunication, CD ROM, Laser printers, signal processing, etc. Quantum dot (QD) lasers have been of interest due to their exclusive characteristics such as law threshold current, low temperature sensitivity, and high optical gain, quantum efficiency and modulation speed are superior to other types of semiconductor lasers, owing to their discrete density of states[1],[2].

Until now, many researchers have surveyed the effect of factors such as cavity length, working temperature[3], QD size[4], indium percentage[5], and substrate index[6] on the laser. In this paper, we aim to investigate the effect of gain compression factor on laser outputs in a single level InAs/GaAs laser[7].

The rest of this article is scheduled as follows: firstly, our single level model is explained in section II. In section III the simulation results are represented. We conclude then in section IV.

## II. Our Model

To study the dynamics of our quantum dot laser, we have obtained the rate equations through energy level model illustrated in Fig. (1)[8]

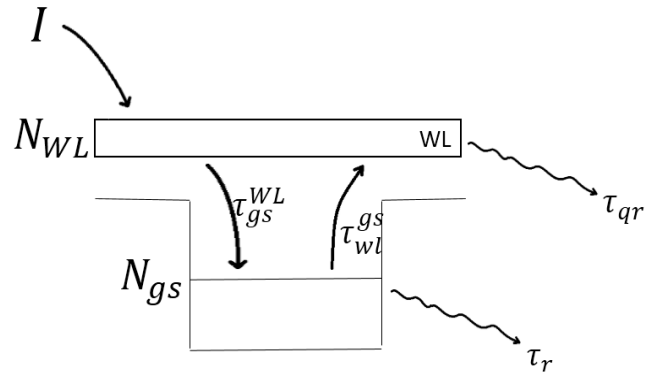

*Fig. 1: The model for relaxation and escape in conduction band of InAs/GaAs quantum dot laser*

- **Rate equations:**

Considering the transitions taken into account in our model, in which only a wetting layer and a ground state are given, three rate equations can be achieved as follows:

$$\frac{dN_{wl}}{dt} = \frac{I}{e} + \frac{N_{gs}}{\tau_{wl}^{gs}} - \frac{N_{wl}}{\tau_r} - \frac{N_{wl}}{\tau_{gs}^{wl}} \quad (1)$$

$$\frac{dN_{gs}}{dt} = \frac{N_{WL}}{\tau_{gs}^{wl}} - \frac{N_{gs}}{\tau_r} - \frac{N_{gs}}{\tau_{wl}^{gs}} - \Gamma V_g K_g (2P_g - 1)\frac{S_{gs}}{1+\varepsilon_{gs}S_{gs}} \quad (2)$$

$$\frac{dS_{gs}}{dt} = \Gamma V_g K_g (2P_g - 1)\frac{S_{gs}}{1+\varepsilon_{gs}S_{gs}} - \frac{S_{gs}}{\tau_p} + \beta\frac{N_{gs}}{\tau_r} \quad (3)$$

Equations (1) to (3) must be solved simultaneously to obtain dynamical behavior of the laser. These equations can give both carrier and photon numbers[9, 10].

In equations (1) to (3), $N_{wl}, N_{gs}$ and $S_{gs}$ are respectively the carrier population in the wetting layer and ground state, and photon population at ground state. Also, $\tau$ denotes the time for a transition detailed in Table 1 and following equations.

$\varepsilon_{gs}$ is the nonlinear gain coefficient for the ground state level obtained as follows:

$$\varepsilon_{gs} = \varepsilon_{mg}\frac{\Gamma}{V_a} \quad (4)$$

where $V_a$ is the active region volume and $\varepsilon_{mg}$ is the gain compression factor defined as:

$$\varepsilon_{mg} = \frac{e^2|P_{cv}|^2 t_p \Gamma}{2n_r^2 m_0^2 \varepsilon_0 E_{gs} V_a \frac{\Gamma_{hom}}{2\hbar}} \quad (5)$$

Here, $\Gamma$ is the optical confinement factor, $\Gamma_{hom}$ is the homogeneous broadening, $|P_{cv}|^2$ is the square of transition matrix element, and $\tau_p$ is the photon lifetime.

- **Output power:**

Output power of the ground state is as follows:

$$P_{out,gs} = \frac{cE_{gs}S_{gs}\log\left(\frac{1}{R_1}\right)}{2n_g L_{ca}} \quad (6)$$

in which $L_{ca}$ is the cavity length and $R_1$ is the mirror reflection coefficient.

$P_g$ is the probability of occupation of GS:

$$P_g = \frac{N_{gs}}{D_{gs}N_D} \quad (7)$$

Relaxation and escape time are interrelated by the following expressions:

$$\tau_{wl}^{gs} = \tau_{gs}^{wl}\frac{D_{gs}N_b}{\rho_{wl}}\exp\left(\frac{E_{wl}-E_{gs}}{k_b T}\right) \quad \text{and}$$

$$\tau_{gs}^{wl} = \frac{\tau_{gs0}^{wl}}{1-P_g} \quad (8)$$

in which $\rho_{wl}$ is the effective density of states of WL, obtained by:

$$\rho_{wl} = \frac{m_e k_b T}{\pi\hbar^2} \quad (9)$$

*Table 1: Parameters used in the simulation*

| PARAMETER | VALUE |
|---|---|
| Light speed ($C$) | $3\times10^8$ m/s |
| Recombination in the QD ($t_r$) | $2.8\times10^{-9}$ s |
| Average refractive index ($n_r$) | 3.5 |
| Group velocity ($V_g$) | $8.571\times10^7$ m/s |
| Degeneracy in GS ($D_{gs}$) | 2 |
| WL energy ($E_{wl}$) | 1100 meV |
| GS energy ($E_{gs}$) | 1000 meV |
| Temperature | 293K |
| Initial relaxation time from GS to WL ($\tau_{es0}^{wl}$) | 5 ps |
| Cavity length ($L_{ca}$) | $900\times10^{-6}$ m |
| Mirror reflection ($R_1$) | 0.3 |
| Mirror reflection ($R_2$) | 0.9 |
| Optical confinement factor ($\Gamma$) | 0.06 |
| Inhomogeneous broadening ($\Gamma_{inh}$) | 20 meV |
| Homogeneous broadening ($\Gamma_{hom}$) | 10 meV |
| Boltzmann constant ($k_b$) | $1.381\times10^{-23}$ m |

### III. Simulation results

By solving the rate equations, plots of the temporal evolution of photons number, the number of photon and output power versus current, and small signal modulation response function are investigated for three different values of gain compression factor for GS.

As it is shown in figure 2, a turn-on delay which is in the order of nanoseconds is needed in which carriers are integrated in the level to a saturation value. Then the laser starts to work by some relaxation oscillations, and after a while, laser arrives at its stable state.

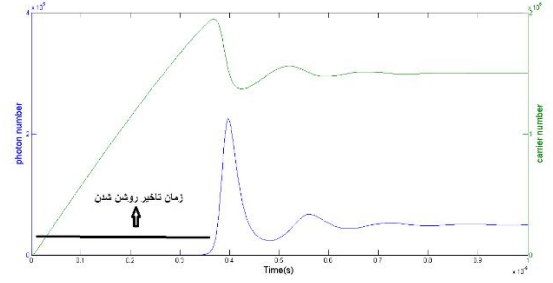

*Fig. 2: Laser Turn-on delay scheme*

As it is seen in Fig. 3, increased gain compression factor leads to reduced number of photons and relaxation oscillations; although it has had no impact on the turn-on delay.

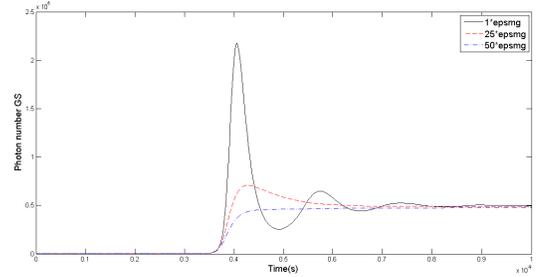

*Fig. 3: Number of photons in time for three values of compression for ground state.*

Fig. 4 indicates that for a given current, number of photons decreases in the ground state by increasing the gain compression factor. However, the threshold current as it sounds remains fixed.

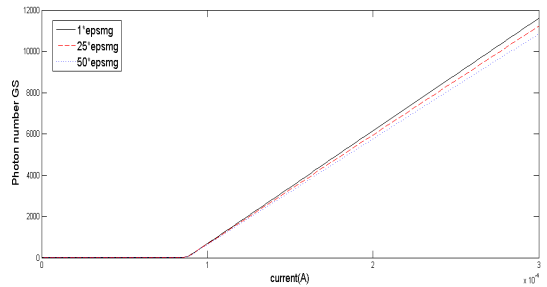

*Fig. 4: Number of photons in terms of current for three different values of compression fro ground state*

Also, in Fig. (5) the modulation response function is plotted as a function of frequency at three different currents. The intersection of the response function with the horizontal axis in -3dB represents the response function bandwidth, and the frequency corresponding to the response function peak is the resonance response function. As it is clear, increasing of the injected current leads to a wider modulation bandwidth and higher resonance frequency.

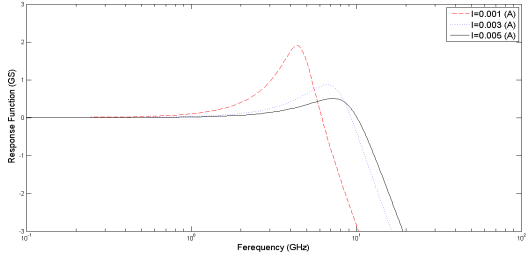

*Fig. 5: Small signal modulation response for three different currents for ground state.*

Fig. (6) depicts the output power versus current for three different values of gain compression factor in the GS. As it is seen, output power is reduced by increasing the gain compression factor[11].

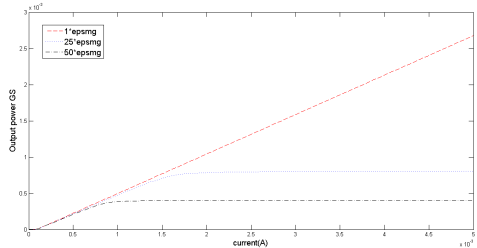

*Fig. 6: Output power versus current for three different compressions for ground state level*

Moreover, in Fig. (7) the modulation response function is plotted as a function of frequency at three different values of gain compression factor. Obviously, increased compression factor leads to decreased value of the modulation bandwidth and resonance frequency in GS. This result coincides with the results found for photon number in time.

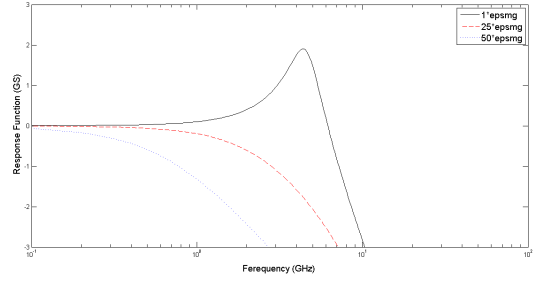

*Fig. 7: Small signal modulation response for three different compressions for ground state.*

### IV. Conclusion

By the above charts we find that enhanced gain compression factor which is the nonlinear term of the optical gain leads to reduced dynamic characteristics of the quantum dot laser. Therefore, this factor must be minimized for optimum characteristics.

### V. Acknowledgment